\documentclass[12pt]{article}
\renewcommand{\i}{\imath}

\newcommand{\bi}{{\bar\i}}

\newcommand{\sect}[1]{\section{#1}\setcounter{equation}{0}}

\newcommand{\OL}[1]{ \hspace{1pt}\overline{\hspace{-1pt}#1
   \hspace{-1pt}}\hspace{1pt} }
\newcommand{\skyp}[1]{}

\begin{document}

\bigskip
\hskip 5in\vbox{\baselineskip12pt
\hbox{NSF-ITP-01-51}
\hbox{hep-th/yymmnnn}}
\bigskip\bigskip

\centerline{\Large \bf Gauge/Gravity Duals with Holomorphic Dilaton
}
\bigskip
\centerline{\bf Mariana Gra\~na}
\medskip
\centerline{Department of Physics}
\centerline{University of California}
\centerline{Santa Barbara, CA 93106}
\centerline{\it mariana@physics.ucsb.edu}
\bigskip
\centerline{\bf Joseph Polchinski}
\medskip
\centerline{Institute for Theoretical Physics}
\centerline{University of California}
\centerline{Santa Barbara, CA\ \ 93106-4030}
\centerline{\it joep@itp.ucsb.edu}

\begin{abstract}

We consider configurations of D7-branes and whole and fractional D3-branes
with ${\cal N}=2$ supersymmetry.
On the supergravity side these have a warp factor, three-form flux and a
nonconstant dilaton. We discuss general IIB solutions of this type
and then obtain the specific solutions for the D7/D3 system.  On the gauge side the
D7-branes add matter in the fundamental representation of the D3-brane gauge
theory. We find that the gauge and supergravity metrics on moduli space agree. 
However, in many cases the supergravity curvature is large even when
the gauge theory is strongly coupled.  In these cases we argue that the useful
supergravity dual must be a IIA configuration.

\end{abstract}

\newpage
\baselineskip=17pt

\sect{Introduction}

The extension of Maldacena's gauge/gravity duality~\cite{maldacena} to systems
with less supersymmetry and richer matter content is an interesting one, both for
understanding more general gauge theories and for application to the local
geometries of warped compactifications~\cite{us}.  A natural extension is to add
D7-branes, as these contribute matter fields in the fundamental representation. 
Thus in this paper we consider ${\cal N}=2$ systems of D7-branes with whole and
fractional D3-branes.

Gauge/gravity duality with many D7-branes has received little consideration.
As far as we are aware, only ref.~\cite{Aharonyetal} directly overlaps our
work, with a discussion of D7-branes and
whole D3-branes.  Ref.~\cite{kehag} allows for a position-dependent dilaton
but requires that it be constant on an AdS$_5$ factor.  There has also been
substantial discussion of configurations of D7-branes and O7-planes such that
the dilaton is everywhere constant, beginning with
refs.~\cite{condil,Aharonyetal}.  The nontrivial dilaton in the present case
brings in new features and puzzles.

In section~2 we review some of the special classes of 
IIB supergravity solution that have played a role in gauge/gravity 
duality and string compactification, and
develop the detailed form of the IIB solutions 
with holomorphic $\tau$. 

In section~3 we find solutions with D7-branes and whole and
fractional D3-branes.  The solutions are singular at long distance, 
but we conjecture that this can be thought of as a UV effect that decouples from
the gauge dual.  In the fractional D3 case the D7-branes are wrapped
on the ALE space ${\bf R}^4/{\bf Z}_2$.  To fix the parameters in the solution we
analyze the induced charges on the D7 world-volume.

In section~4 we first determine the spectrum of the dual
gauge theory and obtain its one-loop effective action.  For D7-branes on ${\bf
R}^4/{\bf Z}_2$ there are two choices of Chan-Paton action, just as for D3-branes
on this space; we relate this to the induced D5 charge.  We then find the
one-loop metric on moduli space and show that it agrees with the action of a
probe in the corresponding dual geometry.

In section~5 we analyze the range of validity of the supergravity duals and find
an unpleasant surprise: even when the gauge theory is strongly coupled, in many
cases the supergravity curvature is large.  This occurs, for example, in the
simple and interesting case of the conformal theory of $SU(N)$ with $2N$
hypermultiplets.  We argue that the correct supergravity dual is instead a IIA
configuration, whose study we leave for future work.

\sect{Solutions: generalities}

\subsection{Special IIB solutions}

Supersymmetric warped solutions of IIB supergravity have recently played an
extensive role in gauge/gravity duality and string compactification.  The
general
solution of this type is not known.  Early papers~\cite{early} obtained very
restrictive results by use of the integrated Bianchi identity for the five-form
flux.  These restrictions need not hold when the transverse dimensions are
noncompact, or when appropriate brane sources are included.\footnote
{Even without supersymmetry, the integrated Einstein equation implies that in a
compact space without branes, warping is impossible in a Minkowski
solution~\cite{hari,MN}.  With appropriate brane sources, or in the
noncompact case, warped solutions are possible; see ref.~\cite{us} for a recent
discussion.}

Much recent work has involved two special cases, which can be characterized
by the
form of the ten-dimensional supersymmetry spinor $\varepsilon$.  This can be
decomposed
\begin{equation}
\varepsilon = \zeta \otimes \chi_1 + \zeta^* \otimes \chi_2^*\ .
\label{decomp}
\end{equation}
Here $\zeta$ is a four-dimensional chiral spinor, $\Gamma^4 \zeta = \zeta$, and
$\chi_{1,2}$ are six-dimensional chiral spinors, $\Gamma^6 \chi_i = -\chi_i$.
Each independent pair $(\chi_1,\chi_2)$ gives rise to one $D=4$ supersymmetry.
The two special cases are then
\begin{eqnarray}
\mbox{Type A(ndy):}&& \chi_2 = e^{i\psi} \chi_1 \ ,\quad \psi \mbox{ real
and constant\ ;} \nonumber\\
\mbox{Type B(ecker):}&& \chi_2 = 0\ \mbox{(or $\chi_1 = 0$)}\ .
\end{eqnarray}
The behavior of the spinor correlates with that of the complex three-form flux
$G_{(3)}$.  In type A solutions $G_{(3)}$  must have a constant phase.  In type 
B
solutions it must be imaginary self-dual; more specifically (see subsection~2.2),
it must be of type (2,1) and primitive with respect to the complex structure of the
transverse space. Vanishing $G_{(3)}$ also gives type B solutions.
Pure brane systems are of one or the other of these types: the D5-brane and
NS5-brane are of type A, and the D3-brane and D7-brane are of type B.

The type A solutions are closely related to the warped heterotic solutions
found by Strominger~\cite{torsion}.  The IIB form was discussed in
ref.~\cite{PapT}. The Chamseddine-Volkov solution~\cite{MN2} is a notable
AdS/CFT example of this type.

The type B solutions are dual to M theory solutions found by Becker and
Becker~\cite{beck^2,GVW,DGS}.  In the M theory form the corresponding
restriction on the supersymmetry spinor is that it have definite
eight-dimensional chirality.
The explicit IIB form was obtained in
refs.~\cite{GP,gubser} for the special case of a constant dilaton.  Such
solutions
have played an important role in gauge/gravity duality.  The ${\cal N} = 1$
fractional brane solution~\cite{KS} is of this form, as well as its ${\cal
N} = 2$
generalization~\cite{n2,n2me}.

In general, the type B solutions allow a holomorphic dilaton.  We find these
solutions in section 2.2.  The various branes in our system --- D7-branes
and whole
and fractional D3-branes --- all preserve supersymmetries of type B.  Moreover,
the supersymmetries preserved by the different branes have a nontrivial
intersection, which is the ${\cal N} = 2$ of the whole system.  Thus these
solutions are the relevant ones.

Finally, we should note that there are interesting solutions which
are of neither special form.  A D3/D5 bound state will interpolate between
type A at short distance and type B at long distance.  Also, the $G_{(3)}$
flux corresponding 
to an ${\cal N} = 1$ or ${\cal N}= 2$ mass perturbation of the ${\cal N}=4$
gauge
theory is of neither type, as one can see from the explicit expressions in
section
III.C of ref.~\cite{PS}.  Full solutions are known only for a few special
states in
the mass-perturbed theory~\cite{special}. In ref.~\cite{PS} an approximate
solution
was found, whose supersymmetry was verified in ref.~\cite{GP}.  This
approximation is
valid over most of parameter space, but it was emphasized that important physics
occurs in regions where it breaks down.

\subsection{Type B solutions}

The solutions of type B could be obtained by duality~\cite{GVW,DGS} from
those of ref.~\cite{beck^2}, but we have found that it is generally simpler
to work
directly in IIB variables.  This section extends the results of
refs.~\cite{GP,gubser}, which were obtained for constant $\tau$.

We first review the relevant results from type IIB supergravity~\cite{Schwarz}.
The massless bosonic fields of the type IIB superstring theory consist of
the
dilaton $\Phi$, the metric tensor $G_{MN}$ and the antisymmetric 2-tensor
$B_{MN}$ in
the NS-NS sector, and the axion $C$, the 2-form potential $C_{MN}$, and the
four-form field $C_{MNPQ}$ with self-dual five-form field strength in the
R-R sector.  Their fermionic superpartners are a complex Weyl
gravitino $\psi_{M}$ ($\Gamma^{10}\psi_{M}=-\psi_{M}$) and a complex Weyl
dilatino $\lambda$ ($\Gamma^{10}\lambda=\lambda$). The theory has $D=10$,
$\mathcal
{N}$=2 supersymmetry with a complex chiral supersymmetry parameter
$\varepsilon$ ($\Gamma^{10}\varepsilon =-\varepsilon$).
The
two scalars can be combined into a complex field $\tau = C+
ie^{-\phi} \equiv \tau_1 + i \tau_2$ which
parameterizes the $SL(2,{\bf R})/ U\left(1\right)$ coset space.

We want to find
backgrounds with four-dimensional Lorentz invariance that preserve some
supersymmetry.  Assuming that the background
fermi fields vanish, we have to find a combination of the bosonic fields
such that the supersymmetry variation of the fermionic fields is zero.
The dilatino and gravitino variations are \cite{Schwarz}
\begin{eqnarray}
\delta\lambda^* &=& -\frac{i}{\kappa} \gamma^{M}P_{M}^*\varepsilon +
\frac{i}{4} G^* \varepsilon^*\ ,
\label{dilatino}
\\
\delta\psi_{M} &=& \frac{1}{\kappa} \Bigl(D_{M} -{i\over2} Q_M \Bigr)
\varepsilon + \frac{i}{480}
\gamma^{M_{1}...M_{5}}
F_{M_{1}...M_{5}}\gamma_{M}
\varepsilon - \frac{1}{16} \Gamma_M G \varepsilon^{*} - \frac{1}{8}
G \Gamma_M \varepsilon^{*}\ .
\label{gravitino}
\end{eqnarray}
Here $G = \frac{1}{6} G_{MNP} \gamma^{MNP}$,
$D_M$ is the covariant
derivative with respect to the metric $g_{MN}$, and
\begin{eqnarray}
P_{M} &=& f^2 \partial_M B\ ,
\quad Q_{M}= f^2 {\rm Im}(B \partial_M B^*) \ ,\label{PQ}
\nonumber\\[2pt]
B &=& \frac{1+i\tau}{1-i\tau}
\ ,\quad \tau = C + i e^{-\Phi}\ ,\quad
f^{-2} = 1 - BB^*
\ .
\end{eqnarray}
The field strengths are
\begin{eqnarray}
G_{(3)} &=& f(F_{(3)}-BF_{(3)}^*)\ ,\quad
F_{(3)} = dA_{(2)} \ ,\nonumber\\
F_{(5)}&=& dA_{(4)}- \frac{\kappa}{8}\,{\rm Im}( A_{(2)}\wedge
F_{(3)}^*)\ .
\label{defpotth}
\end{eqnarray}
with $A_{(2)}$ complex and $A_{(4)}$ real.

We should note that the conventions used in supergravity are different from
those
usually used in string/brane actions, so for reference we give the relations.
The complex potential is related to the NS-NS and R-R potentials by
\begin{equation}
\kappa A_{(2)} = g( B_{(2)} + i C_{(2)} )\ ,
\end{equation}
and the associated fluxes are related by
\begin{equation}
\kappa G_{(3)} = i g e^{i\theta} \frac{F_{(3)\rm s} - \tau
H_{(3)\rm s}}{\sqrt{\tau_2}} \ ,\quad e^{i\theta} =
\biggl( \frac{1 + i \tau^*}{1 - i \tau}  \biggr)^{1/2}
\label{G3}
\end{equation}
and
\begin{equation}
4\kappa F_{(5)} = g F_{(5)\rm s}\ .
\end{equation}
The subscript `s' denotes the usual string quantities, e.g. the R-R flux is
$F_{(3)\rm s} = d C_{(2)}$, the NS-NS flux is $H_{(3)\rm s} = d B_{(2)}$,
and the
five-form flux is $F_{(5)\rm s} = dC_{(4)} +$Chern-Simons.

Define also
\begin{equation}
G_{(3)\rm s} = F_{(3)\rm s} - \tau H_{(3)\rm s} = -i e^{-i\theta}
\frac{\kappa}{g}
\sqrt{\tau_2} G_{(3)}\ .
\label{G3s}
\end{equation}
Note that supergravity equations are usually written in terms of $\kappa$ and
string/brane equations in terms of $g$, but these are related
\begin{equation}
2\kappa^2 = (2 \pi)^7 g^2 \alpha'^4 \ .
\end{equation}

The general Einstein metric and five-form background with four-dimensional
Poincar\'e invariance is
\begin{eqnarray}
ds^{2} &=& Z^{-1/2}\eta_{\mu\nu}dx^{\mu}dx^{\nu}+ Z^{1/2}
\widetilde{ds}{}_6^2\ ,
\label{metricZ}\\
F_{0123
m} &=& \partial_m h\ .
\label{F5}
\end{eqnarray}
We use subindices $M,N,...=0,...,9$;
$\mu,\nu = 0,1,2,3$; and $m,n,...=4,..,9$.  The warp factor $Z$, the
potential $h \equiv C_{0123}$, and the dilaton-axion $\tau$ depend only on
the transverse
$x^m$. The factor of $Z^{1/2}$ is included in the definition of the
transverse metric for convenience.

For solutions of type B,
\begin{equation}
\varepsilon = \zeta \otimes \chi_1\ ,
\end{equation}
the terms proportional to $\varepsilon$ and
$\varepsilon^*$ in the SUSY variations are linearly independent and so must
vanish separately. Equivalently, the terms independent of $G_{(3)}$ and
those containing
$G_{(3)}$ must vanish separately.  Let us start with the former.

First,
\begin{equation}
\delta \psi_\mu = \kappa^{-1} \partial_\mu \varepsilon -
\frac{1}{8}
\gamma_\mu \gamma^m (\kappa^{-1} \partial_m \ln Z - 4Z \Gamma^4 \partial_m h)
\varepsilon
\ . \label{dpmu}
\end{equation}
The spin connection is calculated for tangent space axes $\hat M$
parallel to the Cartesian coordinate axes $M$.
The Poincar\'e supersymmetries are independent of $x^\mu$ and
and so the vanishing of $\delta \psi_\mu$ implies that
\begin{equation}
h = -\frac{1}{4 \kappa Z}\ .
\end{equation}
The variation of $\psi_m$ now takes the form
\begin{equation}
\kappa \delta \psi_m = \biggl(\tilde D_{m} -{i\over2} Q_m \biggr) \varepsilon
+ \frac{1}{8}\varepsilon
\partial_m \ln Z \ , \label{dpm}
\end{equation}
where $\tilde D_m$ is the covariant derivative for $\widetilde{ds}{}_6^2$.
Thus,
\begin{equation}
\tilde \chi_1 = Z^{1/8} \chi_1
\label{chi}
\end{equation}
is covariantly constant,
\begin{equation}
\biggl(\tilde D_{m} -{i\over2} Q_m \biggr) \tilde\chi_1 = 0 \ . \label{cocon}
\end{equation}

The
connection $\tilde D_m$ is therefore in $U(3)$ and so $\widetilde{ds}{}_6^2$
is complex and K\"ahler.  As in Calabi-Yau compactification, if the first
Chern class of $\tilde D_{m} -{i\over2} Q_m$ vanishes for a given metric,
then there is a metric with the same K\"ahler class and complex structure
such that a covariantly constant $\tilde\chi_1$ exists.
We introduce complex coordinates $z^i$, where
\begin{equation}
\gamma^{\bar\imath} \chi_1 = 0\ ;
\end{equation}
acting on $\chi_1$ with $\gamma^i$, $\gamma^{ij}$, and $\gamma^{ijk}$
generate independent spinors.
The final variation proportional to $\varepsilon$ is that of the dilatino,
whose vanishing implies
\begin{equation}
\gamma^{M}P_{M}^*\chi_1 = \gamma^{i}P_{\bar \imath}^*\chi_1 = 0\ .
\label{dilholo}
\end{equation}
It follows that $B$, and so $\tau$, is holomorphic.

The vanishing of the $\varepsilon^*$ variations now implies
\begin{equation}
G \chi_1 = G \chi_1^* = G \gamma^{\bar \imath} \chi_1^* = 0\ .
\label{conditions}
\end{equation}
Expanding these in term of the independent spinors gives
\begin{equation}
G_{ijk} = G_{ij}\!^j = G_{\bar\imath\bar\jmath\bar k} =
G_{\bar\imath\bar\jmath k} = 0\ . \label{21}
\end{equation}
In other words, $G_{(3)}$ is of type (2,1) and primitive, just as for a constant
dilaton.

In addition the Bianchi identities must be satisfied.  For the three-form flux
these are simply
\begin{equation}
dF_{(3)} = dH_{(3)} = 0 \ .
\end{equation}
These of course translate into more complicated identities for $G_{(3)}$ or
$G_{(3)\rm s}$. The five-form flux Bianchi identity implies that
\begin{equation}
- \widetilde\nabla^2 Z = (4\pi)^{1/2} \kappa \rho_3 + \frac{\kappa^2}{12}
G_{pqr} G^{\widetilde{pqr}*} \ .
\label{3den}
\end{equation}

\skyp{
We must now consider the compatibility of this constraint with the Bianchi
identity and with flux quantization.  Let us first consider the case of
constant $\tau$, where the Bianchi identity is simply that
$G{(3)}$ be closed; we are expanding upon remarks made in ref.~\cite{GP}.

For a K\"ahler manifold, the Laplacian takes
$(p,q)$ forms into $(p,q)$ forms, so the harmonic forms are of definite
type.  Harmonic forms are closed, so taking $G_{(3)}$ to be any harmonic
$(2,1)$ form satisfies the Bianchi identity.  The primitivity condition
$G_{ij}\!^j = 0$ is the vanishing of a harmonic $(1,0)$ form.  On a
compact Calabi-Yau manifold there are no such forms; more generally this
might impose a finite number of additional conditions.

The R-R $F_{(3)}$ and NS-NS $H_{(3)}$ are also harmonic, but are real and
quantized.  In general they are then linear combinations of three-forms of
all types, $(3,0)$, $(2,1)$, $(1,2)$, and $(0,3)$.  The condition that
$G_{(3)}$ be $(2,1)$ then gives a finite number of equations for the
coefficients (whereas in principle eq.~(\ref{21}) is an infinite number of
equations, since it must hold at every point), so the quantization, SUSY,
and Bianchi conditions will generically hold on a submanifold of moduli
space.
}

\sect{Solutions with D7-branes}

\subsection{D7+D3-branes}

As a warmup we consider D7-branes and D3-branes in a flat background, rederiving
results obtained in ref.~\cite{Aharonyetal}.  The D3-branes are extended
along the
$\mu$-directions, and D7-branes along the noncompact $\mu$-directions as well
as the 4567-directions.

From the discussion in the previous section, we can take any solution without
D3-branes ($Z=1$, implying $F_{(5)}=0$) and introduce D3-branes through a
nontrivial
$Z$.  Thus we describe first the D7-branes~\cite{stringycstrings}.  We will
use the complex coordinates
\begin{equation}
z^1=\frac{x^4+ix^5}{\sqrt2}\ ,\quad
z^2=\frac{x^6+ix^7}{\sqrt2}\ ,
\quad z=\frac{x^8+ix^9}{\sqrt2}\ .
\label{zdef}
\end{equation}
The dilaton $\tau$ must be holomorphic, and in the given configuration it
depends
only on $z$.
The transverse metric is of the form
\begin{equation}
\widetilde{ds}{}_6^{2}=2\Bigl(
dz^{1}d\bar{z}^{1}+ dz^{2}d\bar{z}^{2}+ e^{\psi(z,\bar{z})}
dz \,d\bar{z} \Bigr)
\label{metric}
\end{equation}
where $\psi$ is to be determined in terms of the dilaton.

Now consider the supersymmetry of this solution.  For arbitrary holomorphic
$\tau(z)$, the covariant constancy condition~(\ref{cocon}) becomes
\begin{eqnarray}
&&\biggl(\partial_{i} +
\frac{1}{4} \tilde{\omega}_{i} ^{ab}
\Gamma_{ab}-\frac{1}{4}\partial_{i} \ln(1-BB^*)\biggr)\tilde{\chi}_1=0\ ,
\nonumber\\
&&\biggl(\partial_{\bi} +
\frac{1}{4} \tilde{\omega}_{\bi} ^{ab}
\Gamma_{ab}+\frac{1}{4}\partial_{\bi} \ln(1-BB^*)\biggr)\tilde{\chi}_1=0\ ,
\label{susyvar}
\end{eqnarray}
where $\tilde{\omega}_{i} ^{ab}$ is the Christoffel connection for the tilded
metric.  For the metric~(\ref{metric}) these become $\partial_{1}
\tilde{\chi}_1=\partial_{2}
\tilde{\chi}_1=\partial_{\bar 1}
\tilde{\chi}_1=\partial_{\bar 2}
\tilde{\chi}_1= 0$
and
\begin{eqnarray}
\partial_{z} \tilde{\chi}_1 &=& +
\frac{1}{4} \tilde{\chi}_1\, \partial_{z}\Bigl[\psi-\ln(1-BB^*)\Bigr]\ ,
\nonumber\\
\partial_{\bar z} \tilde{\chi}_1 &=&
-\frac{1}{4} \tilde{\chi}_1\, \partial_{\bar z}\Bigl[\psi-\ln(1-BB^*)\Bigr]
\ .
\end{eqnarray}
These are integrable provided
\begin{equation}
\psi-\ln(1-BB^*) = \gamma + \gamma^*
\end{equation}
for arbitrary holomorphic $\gamma(z)$.  Then $\tilde{\chi}_1 = e^{4(\gamma -
\gamma^*)}
\eta_0$, where $\eta_0$ is a constant spinor satisfying
$\gamma^{\bar\imath} \eta_0
= 0$, and
\begin{equation}
e^{\psi}= (1-BB^*)e^{\gamma + \gamma^*} \ .
\label{psi}
\end{equation}
Noting that
$1-BB^*={4\tau_2}/{|1-i \tau|^2}$, the holomorphic part is determined by modular
invariance~\cite{stringycstrings},
\begin{equation}
e^{\psi}= \tau_2 |\eta(\tau)|^4
\prod_{i=1}^{N_7} \, | {z-z_i} |^{-1/6}\ ,
\label{psimod}
\end{equation}
where $\eta$ is the Dedekind eta function and $z_i$ are the positions
of the D7 branes.  (To avoid clutter we have introduced dimensionless
coordinates; to convert to the coordinates previously defined substitute $z \to
z/r_0$ where
$r_0$ is some fixed reference distance.)

For the purpose of the gauge/gravity duality we are interested in the local
physics
near $N_7$ D7-branes.  In this limit
\begin{equation}
\tau = \frac{i}{g} + \frac{1}{2\pi i} \sum_{i=1}^{N_7} \ln(z-z_i)
\label{dtau} \
,
\end{equation}
with $z, z_i \ll 1$.  The constant could be absorbed into the argument of the
logarithm, but it is convenient to keep it explicit.
When $z, z_i \ll 1$ then $\tau_2 \gg 1$ and the metric simplifies to
\begin{equation}
e^{\psi} = \tau_2 = \frac{1}{g} - \frac{1}{2\pi} \sum_{i=1}^{N_7} \ln|z-z_i| \
.
\label{dpsi}
\end{equation}

Notice that we are contemplating an arbitrarily large number of D7-branes.  The
local form~(\ref{dtau},\,\ref{dpsi}) becomes singular at $z \sim 1$, where
$\tau_2$
goes through zero.  For $N_7 \leq 24$, this  local solution can be
extended to a nonsingular global solution~(\ref{psimod}).  For $N_7 > 24$ there
is no known nonsingular extension.  Nevertheless, we will use this local
solution and find that it gives sensible results at small $z$.  It is an
interesting question for the future, whether there is any physical realization
of $N_7 > 24$ in string theory, and if not whether the use of this local
solution
in gauge/gravity duality is nonetheless justified.

The D3/D7 solution is now obtained by including a nontrivial $Z$.  This is
determined by the Bianchi identity:
\begin{equation}
- 2\Bigl(e^{\psi} \partial_1
\partial_{\bar{1}}+e^{\psi} \partial_2 \partial_{\bar{2}}+
\partial_3 \partial_{\bar{3}} \Bigr) Z =
(4\pi)^{1/2} \kappa e^{\psi} \rho_3\ ,\quad e^{\psi} \rho_3 =
\sum_{j=1}^N
 \delta^6(x^m - x^m_j) \ .
\label{eqZsources}
\end{equation}
As is well-known~\cite{intersect}, this cannot be solved exactly.  In
section~4 we
will discuss some approximate features.

Finally, let us ask for all supersymmetries of this solution.  For a more
general
spinor $\varepsilon'$, the necessary conditions are first the vanishing of
$\delta\psi_\mu$, eq.~(\ref{dpmu}), which implies that $\Gamma^4 \varepsilon' = 
0$
or $\varepsilon' = \zeta \otimes \chi'_1$.  The vanishing of
$\delta\psi_m$, eq.~(\ref{dpm}), then implies
\begin{equation}
\biggl(\tilde D_{m} -{i\over2} Q_m \biggr) \varepsilon
+ \frac{1}{8}\varepsilon
\partial_m \ln Z =0\ ,
\end{equation}
so that $\chi'_1 = Z^{-1/8} e^{4(\gamma - \gamma^*)} \eta$ for any constant spinor
$\eta$. Finally, the vanishing of $\delta\lambda$ implies that $\gamma^{\bi} P^*_i
\eta =
0$, and so $\gamma^{\bar 3} P^*_i \eta = 0$.  This has two solutions of the
given
chirality, $\eta = \eta_0$ and $\eta = \gamma^{12} \eta_0$.
We can characterize these as the two spinors having definite chiralities in the
4567- and 89-directions,
\begin{equation}
i^2 \gamma^{4567} \eta = i \gamma^{89} \eta = \eta\ .   \label{chir}
\end{equation}
Thus,
as expected, this background has four complex or eight real
supersymmetries, i.e.
$D=4$, ${\cal N}=2$.

As shown in section~2.2, we can add
three-form flux to the above solution provided that $G_{(3)}$, or equivalently
$G_{(3)\rm s}$,  is $(2,1)$, primitive, and satisfies the appropriate Bianchi
identity.  The simplest solution of this form is
\begin{equation}
G_{(3)\rm s} = \tau_2 g(\bar z^3) dz^1 \wedge dz^2 \wedge d\bar z^3
\end{equation}
for any antiholomorphic $g(\bar z^3)$.
The primitivity and $(2,1)$ properties are evident, and the Bianchi identity can
readily be verified.  For $g = (\bar z^3)^k$, this scales as a dimension~$7+k$
perturbation of the gauge theory~\cite{WGKP},
and so does not affect the infrared physics.

\subsection{D7+fractional D3-branes}

From the study of gauge/gravity duals without D7-branes, we know that
interesting gauge theories are obtained by taking a ${\bf Z}_2$ orbifold and
including fractional D3-branes on the fixed plane, corresponding to D5-branes
wrapped on the collapsed 2-cycle.  These duals are developed in
refs.~\cite{GK-KN}.  The full ${\cal N}=2$ supergravity solutions are given in
refs.~\cite{n2,n2me}.  Our main focus will be to generalize these by the
inclusion of D7-branes.

We begin with a ${\bf Z}_2$ orbifold of the D3/D7 solution.  The ${\bf Z}_2$
reflects the 4567-directions, so that the space transverse to the D3-branes is
$({\bf R}^4/{\bf Z}_2) \times {\bf R}^2$ with the D7-branes filling the ${\bf
R}^4/{\bf Z}_2$ ALE space and at a point in the ${\bf R}^2$ parameterized by
$z$.  This preserves ${\cal N} = 2$ supersymmetry.

The D3/D7 solution survives in the orbifolded
theory, provided that $\rho_3$ and therefore $Z$ are invariant under
the orbifolding, and this solution is our starting point.  Then, as discussed in
section~2.2, we can add three-form flux subject to the appropriate conditions.
The new feature of the orbifolded theory is the existence of a zero-size
two-sphere at the fixed point, which is associated with a harmonic
2-form $\omega_{(2)}$ also localized at the fixed point.
It is a standard
property of ALE spaces that $\omega_{(2)}$ is $(1,1)$ and primitive in the ALE
space.  Then, as in the case without D7-branes (our discussion and notation
follow ref.~\cite{n2me}, except that the signs of $B_{(2)}$ and $C_{(2)}$
are reversed to agree with conventions used elsewhere), we take
\begin{eqnarray}
B_{(2)}= 2\pi\alpha' \theta_{\rm B}(z,\bar{z})\omega_{(2)}\ , \quad
C_{(2)}= 2\pi\alpha' \theta_{\rm C}(z,\bar{z})\omega_{(2)}\ .
\label{pot}
\end{eqnarray}
Conversely, at each fixed $z$,
\begin{equation}
\theta_{\rm B} = \frac{1}{2\pi\alpha'} \int_{S^2} B_{\it 2}\ ,\quad
\theta_{\rm C} = \frac{1}{2\pi\alpha'} \int_{S^2} C_{\it 2}\ .
\end{equation}

The Bianchi identity and primitivity condition are automatic.  The condition
that the $(1,2)$ part of $G_{(3)}$, or equivalently of $G_{(3)\rm s}$, vanish is
then
\begin{equation}
\partial_{\bar z} \theta = 0\ ,\quad \theta =  \theta_{\rm C} - \tau \theta_{\rm
B}\ .
\end{equation}
We have used $d \omega_{(2)} = 0$.  Thus $\theta$ is any holomorphic
function,
\begin{equation}
\theta = \mbox{holomorphic}\ .\label{holoth}
\end{equation}
Writing $\theta(z) = \theta_1 + i \theta_2$, the real and
imaginary parts of $\theta =  \theta_{\rm C} - \tau \theta_{\rm
B}$ imply that
\begin{equation}
\theta_{\rm B} = -\frac{\theta_2}{\tau_2}\ , \quad
\theta_{\rm C} = \theta_1 - \frac{\tau_1 \theta_2}{\tau_2} \ .
\end{equation}

The angles $\theta_{\rm B}$ and $\theta_{\rm C}$ are periodic with period
$2\pi$.  A wrapped D5-brane couples magnetically to $\theta_{\rm C}$
and so $\theta$ has a branch cut
\begin{equation}
\theta \sim \pm 2 i \ln(z - z_{5})\ .
\end{equation}
Here $z_5$ is the D5-brane position, the upper/lower signs refer
to D5/anti-D5, and the factor of 2 arises because the two-sphere has
self-intersection number $2$ (this discussed more generally in
ref.~\cite{GK-KN}).

The combination $\theta$ is invariant under the $SL(2,{\bf
Z})$ monodromy of the D7-branes, but may still have branch cuts at the D7-branes
arising from induced D5 charge.  Consider then the Chern-Simons   action for a
D7-brane, whose relevant terms are
\begin{equation}
S_{\rm CS} = \mu_7 \int_{M^4 \times \rm ALE}
 \biggl\{ C_{(8)} + 2\pi\alpha'{\cal F}_{(2)} \wedge
C_{(6)} + \frac{1}{2}(2\pi\alpha')^2 {\cal F}_{(2)}  \wedge {\cal
F}_{(2)} \wedge C_{(4)} \biggr\}\ ;
\end{equation}
(there is also a curvature term that will be discussed shortly).
Here $2\pi\alpha'{\cal F}_{(2)} = 2\pi\alpha'{ F}_{(2)} - B_{(2)}$.
Using the form~(\ref{pot}) and defining ${ F}_{(2)} = \Phi \omega_{(2)}$,
this becomes
\begin{eqnarray}
S_{\rm CS} &=& \mu_7 \int_{M^4 \times \rm ALE}
 \biggl\{ C_{(8)} + 2\pi\alpha'(\Phi - \theta_{\rm B}){\omega}_{(2)} \wedge
C_{(6)} + \frac{1}{2}(2\pi\alpha')^2 (\Phi - \theta_{\rm B})^2 {\omega}_{(2)}
\wedge {\omega}_{(2)} \wedge C_{(4)} \biggr\} \nonumber\\
&=& \mu_7 \int_{M^4 \times \rm ALE} C_{(8)} +
\frac{2\pi\alpha' \mu_7}{2}
\int_{M^4 \times S^2}
(\Phi - \theta_{\rm
B}) C_{(6)} + \frac{(2\pi\alpha')^2 \mu_7}{4} \int_{M^4}
(\Phi - \theta_{\rm B})^2 C_{(4)}
\ .
\end{eqnarray}
In going from the first line to the second we have used properties that follow
from Poincar\'e duality, specifically
\begin{equation}
\int_{\rm ALE} \omega_{(2)} \wedge \alpha_{(2)}
= \frac{1}{2} \int_{\rm S^2} \alpha_{(2)}\ , \quad
\int_{\rm S^2} \omega_{(2)} = 1\ ,
\end{equation}
for any closed 2-form $\alpha_{(2)}$. 
The $\frac{1}{2}$ again arises from the self-intersection number of the
$S^2$.

Recall that $\mu_5 = (2\pi)^2 \alpha' \mu_7$ and
$\mu_3 = (2\pi)^4 \alpha'^2 \mu_7$~\cite{bigbook}, and that in the orbifold
theory $\theta_{\rm B} = \pi$~\cite{Aspin}.  It follows from the coupling to
$C_{(4)}$ that for $\Phi = 0$, the induced D3 charge is $\frac{1}{16}$.
However, we must also include the curvature terms in the Chern-Simons
action~\cite{BSV}.  These make a contribution~$-\frac{1}{16}$, because on the
space ${\rm K3}=T^4/{\bf Z}_2$ the total induced charge is $-1$.  Thus the net
induced D3 charge on the wrapped D7-brane with $\Phi = 0$ is zero.
Similarly the
induced D5 charge is $-\frac{1}{4}$ times that of a wrapped D5.  For $\Phi =
2\pi$ the induced D3 charge is again zero and the induced D5 charge is
$+\frac{1}{4}$.

These considerations suggest that
\begin{equation}
\theta \stackrel{?}{=} 2 i \sum_{i = 1}^{N_7} q_{5i} \ln(z - z_{7i})
+ 2i  \sum_{j = 1}^{N_5} \ln(z - z_{5j})
- 2i  \sum_{k = 1}^{N_{\bar 5}} \ln(z - z_{\bar 5k})
\ . \label{guess}
\end{equation}
Here $i$ runs over D7-branes, $j$ over D5-branes and $k$ over anti-D5 branes.
The induced charges $q_{5i}$
are $\pm \frac{1}{4}$ from the above discussion.  Recall also that
\begin{equation}
\tau = \frac{i}{g} +
\frac{1}{2\pi i} \sum_{i = 1}^{N_7} \ln(z - z_{7i})\ . \label{tsol}
\end{equation}
The form~(\ref{guess}) is not quite correct, as we must include the explicit
orbifold background $\theta_{\rm B} = \pi$, and so add $-\pi \tau$ to $\theta$.
The final result is
\begin{equation}
\theta {=} -\frac{i\pi}{g} +
2 i \sum_{i = 1}^{N_7} \biggl(q_{5i} + \frac{1}{4}
\biggr)  \ln(z - z_{7i}) + 2i  \sum_{j = 1}^{N_5} \ln(z - z_{5j})
-2i  \sum_{k = 1}^{N_{\bar 5}} \ln(z - z_{\bar 5k})
\ .
\label{true}
\end{equation}
This is correct in any configuration in which all D5 charges cancel locally, as
it then just gives the orbifold background $\theta_{\rm B} = \pi$.  It can
then be verified for other configurations by moving the D5- and D7-branes
around.
Note that in the final result~(\ref{true}) the shifts in $\theta_{\rm C}$ around
the D7-branes are properly quantized (multiples of $2\pi$), whereas that did
not hold for eq.~(\ref{guess}).

The D7/fractional-D3 system has the same ${\cal N}=2$ supersymmetry as the
D7/D3 system.  The orbifolding preserves supersymmetries~(\ref{chir}) of
positive 4567-chirality.  The fractional brane flux is manifestly $(2,1)$ with
respect to the complex structure defined by the spinor $\eta_0$.  It is
also $(2,1)$ with respect to the complex structure defined by $\gamma^{12}
\eta_0$: this is obtained by replacing $z^i \leftrightarrow \bar z^i$ in the
ALE directions, so $\omega_{(2)}$ remains $(1,1)$.

The 3-form flux now acts as an additional effective D3-brane source for the warp
factor $Z$.  The Bianchi identity~(\ref{eqZsources}) becomes
\begin{equation}
- 2\Bigl(e^{\psi} \partial_1
\partial_{\bar{1}}+e^{\psi} \partial_2 \partial_{\bar{2}}+
\partial_3 \partial_{\bar{3}} \Bigr) Z =
(4\pi)^{1/2} \kappa e^{\psi} \rho_3
+ \frac{1}{2} (2\pi\alpha'g)^2
\delta_{\rm FP} |D_z \theta|^2 \ ,
\label{3den2}
\end{equation}
where $D_z\theta = \partial_z \theta_{\rm C} - \tau \partial_z \theta_{\rm B}$.
We have used the fact that $\omega_{pq} \omega^{\widetilde{pq}} =
\delta_{\rm FP}$ is a $\delta$-function at the fixed point of the ALE space.

\subsubsection{Probe actions}

Regarding this as a 4-dimensional system, the D7-brane positions are fixed while
the D5-brane coordinates parameterize a moduli space of
$N_5$ complex dimensions.  The metric on this
space can be computed both on the supergravity side and on the gauge theory
side.
In this section we find the supergravity metric, and in the next we will compare
it with the gauge theory metric.

To that end we consider the action for a probe D5-brane at the fixed plane,
whose moduli space has one complex dimension. The relevant terms in the
probe brane
action are
\begin{equation}
S/\mu_5 =- \int d^6\xi\, e^{-\Phi}\left[-\det(G+2\pi\alpha'{\cal
F}_{(2)})\right]^{\frac{1}{2}}  + \int C_{(6)} + \int 2\pi\alpha'{\cal
F}_{(2)}\wedge C_{(4)}\ ,
\label{action}
\end{equation}
where the metric $G$ in the DBI action is in the string frame. The
determinant splits into $\det G_{\|}\det(G_{ab}+2\pi\alpha'{\cal F}_{ab})$,
where
$\|$ denotes the 0123 directions and $a,b$ label the directions in the
2-cycle. If the probe is slowly moving with velocity $v$ in the complex plane,
then
\begin{equation}
e^{-\Phi}(-\det G_{\|})^{\frac{1}{2}}=
g^{-1} Z^{-1}\Bigl(1-|v|^2 e^{\psi}
Z\Bigr)^{\frac{1}{2}} \approx g^{-1} Z^{-1}-\frac{1}{2}g^{-1}  e^{\psi} |v|^2
\ .
\label{inter}
\end{equation}
We have used $G_{\rm Einstein}=g^{1/2} e^{-\Phi/2}G_{\rm string}$, as well as
the form  (\ref{metricZ}) and (\ref{metric}) for the Einstein metric.
For the other determinant, $\det(G_{ab}+2\pi\alpha'{\cal F}_{ab})=
\det(2\pi\alpha'{\cal F}_{ab})$, since the 2-cycle is in the limit of zero area.
Slightly blowing up this collapsing cycle so that it is a small 2-sphere,
we get:
\begin{equation}
\int_{S^2} \det(2\pi\alpha'{\cal
F}_{ab})^{\frac{1}{2}}= 2\pi \alpha' \Bigl| 2\pi n - \theta_B \Bigr|\ ,
\label{B}
\end{equation}
where $\int_{S^2} F_{ab} = 2\pi n$ is the quantized D-brane gauge flux.
Combining (\ref{inter}) and (\ref{B}), the DBI Lagrangian density in the
noncompact dimensions becomes
\begin{equation}
{\cal L}_{\rm DBI} = - 2\pi \alpha' \frac{\mu_5}{g} \Bigl| 2\pi n - \theta_B
\Bigr| \biggl( Z^{-1} -\frac{1}{2} e^{\psi} |v|^2 \biggr)\ .
\end{equation}

The potential $C_{(6)}$ is obtained from the seven-form field
strength
\begin{eqnarray}
dC_{(6)} &=& - e^{\Phi}
  {*} ( F_{(3)\rm s} - C H_{(3)\rm s})
+ C_{(4)} \wedge H_{(3)\rm s} \ .
\label{C6}
\end{eqnarray}
The exterior derivative of the right-hand side vanishes by the IIB
supergravity equations; in fact, this consistency condition determines the
form of the Chern-Simons terms here.  Inserting the type B form
for the metric and $C_{(4)}$, the right-hand side is proportional to
Re$(\tilde{*}_6 G_{(3)\rm s} - i G_{(3)\rm s})$, where $\tilde{*}_6$ denotes
the dual in the transverse directions using the metric
$\widetilde{ds}{}^2$.  This combination vanishes as a consequence of the
supersymmetry conditions, so for all type B solutions the coupling to $C_{(6)}$ is
at most a constant in the action, which can be ignored.

The Chern-Simons term, for a type B background, gives the Lagrangian
density
\begin{equation}
{\cal L}_{\rm CS} = 2\pi \alpha' \frac{\mu_5}{gZ} ( 2\pi n - \theta_B
)\ .
\end{equation}
As long as the induced D3 charge
\begin{equation}
q_3 = n - \frac{1}{2\pi} \theta_B
\label{q3}
\end{equation}
is positive, this cancels the potential from the DBI action. The final result 
for the DBI Lagrangian density is
\begin{equation}
{\cal L}_{\rm DBI} = \frac{1}{2} T_3 q_3 e^{\psi} {|v|^2}
\end{equation}
with $T_3 = 4\pi^2 \alpha' \mu_5 / g$ being the D3-brane tension.

This action has a
simple interpretation: the inertial mass
comes entirely from the induced D3-brane charge, with an additional factor of
$e^\psi$ from the effect of the D7-branes on the metric.  For anti-D5-branes
the same result holds with $q_3$ replaced by $n + \frac{1}{2\pi}
\theta_B$.  Recall that these branes are the correct degrees of
freedom on the moduli space only for
\begin{equation}
0 \leq q_3 \leq 1\ . \label{range}
\end{equation}
Where the induced
D3 charge becomes negative the 5-branes would no longer be BPS.  The moduli
space of 5-brane coordinates thus does not continue on into such a region but
rather is joined onto the internal moduli space of the enhanced symmetry
region defined by the curve where $q_3$ vanishes~\cite{enhan}.
When the magnitude of the induced D3 charge exceeds unity, the 5-brane
acquires additional moduli and can separate into elementary constituents in
the range~(\ref{range})~\cite{n2me}.

Carrying out a similar expansion for the gauge field action yields
an additional term
\begin{equation}
{\cal L}_{\rm DBI} = T_3 q_3 \biggl\{ \frac{1}{2} e^{\psi}{|v|^2}
- \frac{1}{4}  (2\pi\alpha')^2 e^{-\Phi}{F_{\mu\nu} F^{\mu\nu} \biggr\}\ .}
\end{equation}
Finally, noting that $e^\psi = e^{-\Phi} = \tau_2$ in our solution, this
becomes
\begin{equation}
{\cal L}_{\rm DBI} = \biggl( n \tau_2 \pm \frac{\theta_2}{2\pi} \biggr)
\biggl(
\frac{1}{2} T_3 {|v|^2}  - \frac{1}{8\pi} F_{\mu\nu} F^{\mu\nu}
\biggr)\ .
\end{equation}
where the plus (minus) corresponds to a D5 (anti-D5) brane. The kinetic and gauge 
terms have the same coefficient, implying that $z$ is
the ${\cal N}=2$ special coordinate.  More generally, as in the F theory
solution~(\ref{psimod}), $e^\psi/e^{-\Phi}$ is the modulus of a holomorphic
function and the special coordinate is then a holomorphic function of $z$.
For future reference we define  $\tau_{2,\rm eff}$ to be the coefficient of
$- \frac{1}{8\pi} F_{\mu\nu} F^{\mu\nu}$, hence
\begin{equation}
\tau_{2,\rm eff} = n \tau_2 \pm \frac{\theta_2}{2\pi}\ . \label{teff}
\end{equation}

\sect{Gauge theory duals}

\subsection{The D7/fractional D3 spectrum}

The gauge theory dual to our supergravity solution is obtained from the
open-string spectrum for D3- and D7-branes on the orbifold~\cite{KachSil,
LawNekVaf}.  The ${\bf
Z}_2$ reflection $R$
acts on the D3 and D7 Chan-Paton degrees of freedom via matrices which in a
diagonal basis will be of the form
\begin{eqnarray}
\gamma_{R3}=\pmatrix{I_{N_{3^+}} & 0 \cr 0 & -I_{N_{3^-}}\cr}\ ,
\quad
\gamma_{R7}=\pmatrix{I_{N_{7^+}} & 0 \cr 0 & -I_{N_{7^-}}\cr}\ ,
\end{eqnarray}
where $I_N$ is the $N\times N$ identity matrix.
The interpretation of $\gamma_{R3}$ is well-known~\cite{fract}.  This basis
represents half D3-branes trapped on the fixed plane.  Geometrically, the
positive eigenvalues correspond to wrapped D5-branes on the collapsed $S^2$,
and the negative eigenvalues to wrapped anti-D5-branes.  Thus,
\begin{equation}
N_{3^+} = N_{5}\ ,\quad N_{3^-} = N_{\bar 5}\ .
\end{equation}
Each D5 carries
one-half unit of D3 charge in the orbifold theory, so the D3 charge
is one-half the number of D3 Chan-Paton indices (this is evident in a
basis in which each D3-brane has an image); thus $Q_3 = \frac{1}{2}
N_3 = \frac{1}{2} (N_{3^+} + N_{3^-})$.

We must similarly deduce the meaning of $\gamma_{R7}$.  There is a natural
guess, since we have seen in section~3.2 that the D7-brane has two
ground states, with D5-charges $\pm \frac{1}{4}$.  Indeed, one can argue for
this connection as follows.\footnote{We thank M. Douglas for suggesting this.}
The reflection $R$ relates opposite points on a given D7-brane, so
$\gamma_{R7}$ represents a phase under a closed motion on ${\bf R}^4/{\bf
Z}_2$. This phase is a D7 Wilson line around the fixed point and so should
arise from a localized flux, which is just the degree of freedom
distinguishing the two D7 states.  To be precise, a disk bounded by the given
closed motion intersects the collapsed $S^2$ once, so the integral of the
flux on this disk is one-half of its integral on the collapsed $S^2$, giving
a phase difference of $\pi$ between the two states.  

In fact, the induced charge
has already been calculated in ref.~\cite{gimpol}, in
$T$-dual 5-9 form, where the last line of eq.~3.30 shows that the induced
charge carried by the D9-brane is $-\frac{1}{4}$ of that carried by the
D5-brane.  So just as for D3-branes, the
D7 Chan-Paton eigenvalue is related to the brane's D5 charge, though with a
different proportionality: Chan-Paton eigenvalue $\pm 1$ corresponds to charge
$\mp \frac{1}{4}$.

The dynamical fields in $D=4$ are obtained from the 3-3, 3-7 and 7-3 strings.
The massless 3-3 spectrum is well-known to be a $U(N_{3^+}) \times
U(N_{3^-})$ gauge theory with two $({ N}_{3^+},\OL{{
N}}_{3^-})\oplus(\OL{{ N}}_{3^+}, { N}_{3^-})$
hypermultiplets~\cite{quiver,GK-KN}.  The action of the  orbifold on
the 3-7 strings is:
\begin{equation}
R|\psi,i,j
\rangle=\gamma_{R3,i i'}\gamma_{R7,jj'}|R\psi,i',j'\rangle\ .
\end{equation}
where $\psi$ is the oscillator state and $i$ and $j$ are the D3 and D7
Chan-Paton indices.
In the Ramond sector, the fermionic zero-modes on the 3-7 strings come from
the 23- and 89- planes, so that the massless fermionic states are
labeled by the  corresponding helicities $|s_1,s_4\rangle$ and the GSO
projection sets $s_1=-s_4$.  The reflection in the 4567-directions has no
action on this state, so the orbifold projection amounts to
$\gamma_{R3,i i}\gamma_{R7,jj} = 1$.  Thus the 3-7 strings contribute
$N_{7^+}$ Weyl fermions of each chirality (from $s_{4} = \pm \frac12$) in
the fundamental $(N_{3^+},1)$ and $N_{7^-}$ Weyl
fermions of each chirality in the
$(1,N_{3^-})$.  The 7-3 strings contribute the antiparticles of
these.  In the NS sector,  states are labeled by the 4567 helicities
$|s_2,s_3\rangle$ and the GSO projection sets $s_2=s_3$.
Supersymmetry requires bosonic partners for the fermions in the
spectrum, so
$R$ must act trivially on the oscillator part of these bosonic states.  This
is so if $R$ is defined as $e^{i\pi(s_2-s_3)}$: this is the condition that
the orbifold and D7/D3 supersymmetries be compatible.

In summary, the massless spectrum is an ${\cal N}=2$ gauge theory with
\begin{eqnarray}
\mbox{vector multiplets}:&&U(N_{3^+}) \times U(N_{3^-})\mbox{ adjoint\
,}
\nonumber\\
2\mbox{ hypermultiplets}:&&(N_{3^+},\OL{N}_{3^-})\oplus
(\OL{N}_{3^+}, { N}_{3^-})\ ,
\nonumber\\
N_{7^+}\mbox{ hypermultiplets}:&&(N_{3^+},1)\oplus
(\OL{N}_{3^+},1)\ ,
\nonumber\\
N_{7^-}\mbox{ hypermultiplets}:&&(1,{N}_{3^-})\oplus
(1, \OL{ N}_{3^-})
\ . \label{spectrum}
\end{eqnarray}
Again, the superscripts $\pm$ refer to the action of the ${\bf Z}_2$ on the D3
and D7 Chan-Paton factors.

\subsection{The metric on moduli space}

The Coulomb branch of moduli space is defined classically by the eigenvalues of
the vector multiplet scalars $\phi$ and $\tilde\phi$.
\begin{equation}
\phi = {\rm diag}(a_1,\ldots,a_{N_{3^+}})\ ,\quad
\tilde\phi = {\rm diag}(\tilde a_1,\ldots,\tilde a_{N_{3^-}})\ .
\end{equation}
These are related to the positions of the fractional branes by
\begin{equation}
{2\pi\alpha'} a_i = {z_{5i}} \ ,\quad {2\pi\alpha'} \tilde a_i = {z_{\bar
5i}} \ .
\end{equation}
 Define similarly for the D7-brane positions
\begin{equation}
{2\pi\alpha'} b_i = {z_{7^+i}} \ ,\quad {2\pi\alpha'} \tilde b_i =
{z_{7^-i}} \ .
\end{equation}

The moduli space metric is obtained from the
$\mathcal N$=2 prepotential $\mathcal F$, whose perturbative form is
\begin{eqnarray}
{\mathcal F} &=& {\mathcal F}_{\rm classical} + {\mathcal F}_{\rm one\; loop}
\ ,
\nonumber\\
{\mathcal F}_{\rm classical} &=& \frac{2\pi i}{g^2_+}
\sum_{i=1}^{N_{3^+}} a_i^2
+
\frac{2\pi i}{g^2_-} \sum_{i=1}^{N_{3^-}} b_i^2 \ ,
\\
{\mathcal F}_{\rm one\; loop} &=&
\frac{i}{8\pi}\Biggl\{ \sum_{i , j =1}^{N_{3^+}}
(a_i-a_j)^2
\ln \frac{(a_i-a_j)^2}{\mu^2}
+ \sum_{i, j=0}^{N_{3^-}}(\tilde a_i-\tilde a_j)^2
\ln \frac{(\tilde a_i-\tilde a_j)^2}{\mu^2}
- \sum_{\rm hypers} m^2 \ln \frac{m^2}{\mu^2} \Biggr\}\ .\nonumber
\end{eqnarray}
Here $g^2_\pm$ are the two classical gauge couplings, each equal to
$8\pi g$ in the classical limit.
The masses of the hypermultiplets in eq.~(\ref{spectrum}) are respectively
\begin{equation}
a_i - \tilde a_j\ , \quad a_i - b_j\ ,\quad \tilde a_i - \tilde b_j
\ .
\end{equation}

The effective value of $\tau_2$, normalized as in eq.~(\ref{teff}),
is Im(${\cal F}''$), and represents  the inverse of the effective
coupling-squared for  the
$U(N_{3^+})$ factor.  To obtain the effective action for a D5-brane probe,
increase the rank of the gauge group by one, adding in the field $a_0$ and
extending the ranges in the sums. Then
\begin{eqnarray}
\tau_{2,\rm eff} &=& {\rm Im}\biggl( \frac{\partial^2 {\cal F}}{\partial a_0^2}
\biggr) \nonumber\\
&=& \frac{1}{2g} + \frac{1}{2\pi} \sum_{i=1}^{N_{3^+}}
\ln \left|\frac{(a_0-a_i)^2}{\mu^2}\right| - \frac{1}{2\pi} \sum_{i=1}^{N_{3^-}}
\ln \left|\frac{(a_0-\tilde a_i)^2}{\mu^2}\right|  - \frac{1}{4\pi} 
\sum_{i=1}^{N_{7^+}}
\ln \left|\frac{(a_0-b_i)^2}{\mu^2}\right|\ .\qquad
\end{eqnarray}
(An uninteresting numerical constant has been absorbed into the definition of
$\mu$.)

The moduli space is divided into regions which are separated by enhan\c con
curves~\cite{enhan,n2me}.  Within each such region the supergravity calculation is
supposed to match the appropriate perturbative description~\cite{enhan2}, with a
nonperturbative rearrangement of degrees of freedom when an enhan\c con is
crossed.  The perturbative orbifold corresponds to the range $0 < 
\frac{\theta_{\rm
B}}{2\pi} < 1$, where eqs.~(\ref{range}) and (\ref{q3}) imply that the D5-brane 
probe corresponds to
$n=1$.  Then eq.~(\ref{teff}) gives
\begin{equation}
\tau_{2,\rm eff} = \tau_2 + \frac{1}{2\pi}
\theta_2\ . \label{teff2}
\end{equation}
Inserting the results~(\ref{tsol},\,\ref{true}) for $\tau$ and
$\theta$, one finds that the metrics do agree, where we identify
$\mu = r_0/2\pi\alpha'$ ($r_0$ is the reference scale introduced below
eq.~(\ref{psimod})).
The metric for an anti-D5 probe also agrees.

\sect{Discussion}

Now let us consider the conditions under which the supergravity solution
gives a good description of the theory.  We first summarize the solution:
\begin{eqnarray}
ds^{2} &=& Z^{-1/2}\eta_{\mu\nu}dx^{\mu}dx^{\nu}+ Z^{1/2}
\widetilde{ds}{}_6^2\ ,\nonumber\\
\widetilde{ds}{}_6^{2} &=& 2\Bigl(
dz^{1}d\bar{z}^{1}+ dz^{2}d\bar{z}^{2}+ \tau_2
dz \,d\bar{z} \Bigr)\ , \nonumber\\
\tau &=& \frac{i}{g} +
\frac{1}{2\pi i} \sum_{i = 1}^{N_7} \ln(z - z_{7i})\ ,\nonumber\\
&&\hspace{-50pt}
- 2\Bigl( \partial_1
\partial_{\bar{1}}+ \partial_2 \partial_{\bar{2}}+ \tau_2^{-1}
\partial_3 \partial_{\bar{3}} \Bigr) Z =
(4\pi)^{1/2} \kappa \rho_3
+ \frac{1}{2\tau_2} (2\pi\alpha'g)^2
\delta_{\rm FP} |D_z \theta|^2 \ ,
\nonumber\\
\theta &=& -\frac{i\pi}{g} +
i \sum_{i = 1}^{N_{7^-}} 
  \ln(z - z_{7^-i}) + 2i  \sum_{j = 1}^{N_5} \ln(z - z_{5j})
-2i  \sum_{k = 1}^{N_{\bar 5}} \ln(z - z_{\bar 5k})
\ .
\end{eqnarray}
As discussed in section~2.1, there is a
singularity in the dilaton at large radius, when $\ln r \sim 2\pi / g N_7$.
In order that the unknown physics of the singularity decouple we then need
$r \ll e^{2\pi/g N_7}$.  It will be convenient to make the slightly stronger
assumption that
\begin{equation}
r \ll 1 \ .
\end{equation}
This inequality also implies that $\tau_2^{-1} \ll 1$ so that the theory is weakly
coupled.  

In addition for a good supergravity dual the string metric must have curvature whose
inertial components are small in string units~\cite{imsy},
\begin{equation}
\alpha'{\cal R}_{\rm s} \ll 1\ .
\end{equation}
A typical term in the curvature is of order
\begin{equation} 
{\cal R}_{\rm s} \sim G_{\rm s}^{z\bar z} |\partial_z \ln Z|^2
= \biggl( \frac{g}{\tau_2 Z} \biggr)^{1/2} |\partial_z \ln Z|^2 \ .
\label{cond1}
\end{equation}
Although we cannot find the warp factor $Z$ exactly, we can estimate it.  For
convenience we take all branes to be at the origin.  The arguments of the logarithms
are small and so the logarithms are slowly varying. One thus obtains a good
estimate of $Z$ at any position by treating $\tau$ and
$\theta$ as constants.  The unwarped metric is then flat, and $Z$ satisfies an
ordinary Laplace equation, so that the warp factor reduces to that for a D3-brane
system.  Thus,
\begin{equation}
Z\approx \frac{R^4}{ \tilde{r}^4} \ ,\quad
\tilde{r}^2={r_1^2+r_2^2} + {\tau_2} r^2\ ,
\quad R^4=4\pi g Q_3 \ , 
\label{Zapprox}
\end{equation}
where $Q_3$ is the total 3-brane charge
\begin{equation}
Q_3=N_{3^+}\biggl(1-\frac{\theta_B}{2\pi}\biggr)+N_{3^-}\frac{\theta_B}{2\pi} \ .
\end{equation}
Evaluating this on the plane $z_1 =
z_2 = 0$, where it is greatest, we find
\begin{equation}
\alpha'{\cal R}_{\rm s} \sim \sqrt{\tau_2/Q_3}\ .
\end{equation}
This has a simple interpretation: it is the condition for the usual AdS/CFT
duality to be valid~\cite{maldacena}, substituting the running values of $Q_3$ and
$\tau$.  Thus we need
\begin{equation}
Q_3 \gg \tau_2 \quad \Rightarrow \quad r \gg e^{-2\pi Q_3/N_7}\ .\label{sucon1}
\end{equation}
In particular, the supergravity dual is good over a range of scales only if
\begin{equation}
Q_3 \gg N_7\ .  \label{sucon2}
\end{equation}

This result is unfortunate.  For example, an interesting case is to take $N$
D3$^+$-branes and $2N$ D7$^+$-branes all at the origin: this gives a conformal
${\cal N}=2$ $SU(N)$ theory with fundamental matter.  The solution for this case is
\begin{equation}
\tau = \frac{i}{g} - \frac{iN}{\pi} \ln z\ , \quad
\theta = -\frac{i\pi}{g} + 2iN\ln z\ , \label{consol}
\end{equation}
giving
\begin{equation}
\tau_{2,\rm eff} = \tau_2 + \frac{\theta_2}{2\pi} = \frac{1}{2g}\ , \quad
\frac{\theta_{\rm B}}{2\pi} = \frac{\pi - 2gN \ln r}{2\pi - 2gN \ln r}\ .
\end{equation}
The effective coupling is constant, as is already assured by the general
agreement between the moduli space metrics as calculated in supergravity and the
gauge theory.  Note that we have here a conformal gauge theory even
though the dilaton is nontrivial.  Unfortunately $Q_3 \stackrel{<}{\sim} N_7$
and so we have not found a good dual.  We will return to this issue shortly.

The condition~(\ref{sucon1}) is satisfied over a wide range of scales in the
case of
$N$ D3$^+$-branes and $N_7 \ll N$ D7$^+$-branes.  This gives a ${\cal N}=2$ $SU(N)$
theory with a small amount of matter.  However, there is a large negative
$\beta$-function so the coupling quickly becomes strong, and so there is
interesting dynamics only over a small range of scales with an
enhan\c con~\cite{enhan} in the IR.

The one case where we obtain a useful dual is $N_{3^+} \sim N_{3^-} \gg N_{7^\pm}$.
This gives an approximately conformal $SU(N_{3^+}) \times SU(N_{3^-})$ theory with
bifundamental matter plus a small amount of fundamental matter.

This raises the interesting question: what {\it is} the dual to the conformal
${\cal N}=2$ $SU(N)$ theory with fundamental matter, when the 't Hooft parameter
$gN$ is large?  Let us see why our dual~(\ref{consol}) fails.  At $r \ll 1$,
$\tau_2$ quickly becomes large, so the underlying string theory is {\it weakly}
coupled. The reason that the gauge dynamics remains strongly coupled is that at
the same time $\theta_{\rm B}$ rapidly approaches the value $2\pi$ at which the
ALE space becomes singular.  One could also try to obtain a dual description
by starting with the dual to a product theory with bifundamental
matter~\cite{GK-KN} and taking one gauge coupling to zero while holding the
other fixed.  The limit is again zero string coupling on a singular
ALE space.

The problem of understanding weakly coupled string theory on a
singular ALE space is a familiar one, and it has been argued that the correct
effective description is obtained by $T$-duality on one of the angular directions
of the ALE space, giving a IIA configuration with parallel
NS5-branes~\cite{alesing}.  With the inclusion of D3- and D7-branes on the IIB side
one obtains D4- and D6-branes on the IIA side.  Such configurations have of course
been extensively considered~\cite{mqcd}, and their $T$-duality to the IIB
configurations discussed~\cite{tdual}.  However, thus far they have been applied
only to the moduli space dynamics.  To obtain a complete dual to the large-$N$
gauge theory one needs the full supergravity solution on the IIA side.  There has
been recent progress in this area~\cite{intm5}, and we hope to return to this
point in future work.

\subsection*{Acknowledgments}

We would like to thank I. Bena, O. de Wolfe, M. Douglas, A. Fayyazuddin, and D.
Smith for discussions and communications. This work was supported by National
Science Foundation grants PHY99-07949 and PHY97-22022.

\end{document}